\documentstyle[
aps,prb,amsmath,bm,multicol
]{revtex}
\begin{document}

\draft
\flushbottom

\title{Universal mechanism of discontinuity of commensurate-incommensurate transitions in three-dimensional solids: Strain dependence of soliton self-energy}

\author{A.P. Levanyuk,$^1$ S.A. Minyukov$^2$ and A. Cano$^1$ }
\address{$^1$Departamento de F\'\i sica de la Materia Condensada, C-III,Universidad Aut\'onoma de Madrid, E-28049 Madrid, Spain\\
$^2$Institute of Crystallography, Russian Academy of Sciences,
Leninskii Prospect 59, Moscow 117333, Russia }
\date{ \today}
\maketitle

\begin{abstract}
We show that there exists a universal mechanism of long-range soliton attraction in three-dimensional solids and, therefore, of discontinuity of any commensurate-incommensurate (C-IC) phase transition. This mechanism is due to the strain dependence of the soliton self-energy and specific features of the solid-state elasticity. The role of this mechanism is studied in detail for a class of C-IC transitions where the IC modulation is one-dimensional, the anisotropy in the order parameter space is small, and the symmetry of the systems allows the existence of the Lifshitz invariant. Two other mechanisms of soliton attraction are operative here but the universal mechanism considered in this paper is found to be the most important one in some cases. Comparison with the most extensively studied C-IC transition in $\rm K_2SeO_4$ shows that the experimentally observed thermal anomalies can be understood as a result of the smearing of the theoretically predicted discontinuous transition. 

\end{abstract}


\begin{multicols}{2}

\section{Introduction}
\vspace{-.25cm}
The earliest theories of commensurate-incommensurate (C-IC) transitions exposed a scenario of continuous transformation.\cite{Therories} According to this classical scenario, C-IC transitions occurred because the self-energy of some kind of soliton became negative so that these solitons appeared spontaneously. The interaction between solitons was found to be completely repulsive. In consequence, the C-IC phase transition temperature coincided with the temperature of the sign change of the soliton self-energy, and the soliton density at this temperature was zero.

The physical meaning of these solitons is different for different systems: they can be dislocations, domain walls, Abrikosov vortices, {\it etc.} The main result of the present paper is valid for all of them in three-dimensional solids, and it is deduced in detail considering these solitons as domain walls. 
This type of solitons appears, for instance, in IC dielectrics, IC magnetics, charge-density-wave systems and spin-Pierls compounds. Although the total number of these examples is difficult to estimate, it numbers in the hundreds. What seems to be more clear is that the classical scenario has never been observed in C-IC transitions: the density of domain walls in IC phases is never small experimentally. Even in the so-called ``domain-like regime'' the distances between domain walls are normally comparable with the domain wall widths and in the more frequently observed ``sinusoidal regime'' the domain walls strongly overlap and loose their individuality. 

Partially, this discrepancy between theoretical predictions and observations is purely experimental. It is because IC systems need a tremendously long time to achieve equilibrium. Its relaxation implies creation and arrangement of domain walls. The driving force of this arrangement is the interaction between the walls that, according to the above mentioned theories, decay exponentially with the interwall distance. At the same time the relaxation is expected to be strongly hampered due to pinning of the walls at defects and effects of discreteness of the crystal lattice. In other words, the theories deal with equilibrium states while in experiments one observes, most probably, systems which are quite far from the equilibrium.

It is also natural to suspect that some of the features of the classical scenario are results of implied approximations and neglect of the theories. So one may ask what parts of this scenario will survive within a more complete theory, a question that arose long ago. There were many proposals to explain why the C-IC transitions should be discontinuous. There were too many, in fact, some of them proved to be unjustified. But still there are two survivor effects, in some sense universal, which lead to the discontinuity: (i) the interaction between the order parameter and the elastic strains via the strain dependence of the coefficient of the so-called Lifshitz invariant\cite{Bruce,Golovko} and (ii) the long-range attraction between the walls due to their thermal and quantum fluctuations.\cite{Levanyuk_Laiz.} In this paper we show that a third effect must be taken into account. It is also a coupling between the order parameter and the elastic strains, but this coupling is described by the striction term. Any symmetry allows this coupling independently of the Lifshitz invariant allowance in the corresponding Landau thermodynamic potential. We shall compare the effects of this striction-mediated mechanism and of the two mentioned above. The relative importance of the three mechanisms is found to be different for different classes of C-IC transitions, with the striction-mediated mechanism the most important at least for some of them.

We mention, first of all, that if the symmetries of the two phases are not related the discontinuity of such a transitions is trivial. We suppose, therefore, that the structure of the IC phase can be presented as a space modulation of the structure of the commensurate phase, i.e., the two phases are describable in terms of the order parameter of the (implicit) normal-commensurate transition. Then one has to distinguish between two cases, that were labelled as C-IC transitions of type I and type II in Ref.\cite{Bruce}. If the Lifshitz invariant is absent, IC-C transitions (type II) are clearly discontinuous in the closeness of the Lifshitz point. Michelson\cite{Michelson} was the first who exposed this conclusion, which was confirmed later by many authors (see, e.g., Refs.\cite{Bruce,Ishibashi78}). However, under special conditions it has been predicted that type-II IC phases can obtain domain-like structures with continuous C-IC transitions.\cite{Buzdin} Recently such a behavior has been discussed for so-called spin-Pierls compounds.\cite{Spin} In this paper we shall discuss in detail cases in which the Lifshitz invariant is present and the IC modulation appears along one crystallographic direction. The C-IC transitions in these cases (type I) are well established as continuous transitions if the above mentioned mechanisms are not taken into account.\cite{Sannikov,Golovko_b} Systems that belong to the discussed cases number in the several tens of C-IC transitions of IC dielectrics\cite{IC_phases_in} and there are at least several examples among magnetic\cite{Yzumov} and charge density wave systems.\cite{Coleman} 

The above described classical scenario has been modified historically following the three lines of the efforts clearly outlined in Ref.\cite{Bruce}: (i) overcome the constant-amplitude approximation, (ii) study the effects of fluctuations neglected within the Landau-like approach and (iii) analyze the role of coupling between the order parameter and other variables.

The development of studies along the first line was far from being smooth. Naive attempts to overcome the constant-amplitude approximation concluded that the transition should be discontinuous. However, subsequent studies showed that the performed procedure was incorrect. The full story can be found in Ref.\cite{Sannikov}. The most consistent and thorough efforts to overcome the constant-amplitude approximation were made by Golovko.\cite{Golovko_b,Golovko_a} His conclusion was, in short, that the C-IC transition remains continuous even well beyond the constant-amplitude approximation in all cases he was able to treat exactly.

Along the second line it was noted first that in the two-dimensional case fluctuations of domain walls enhance the repulsive character of their interaction,\cite{Pokrovsky} while these fluctuations are ineffective in the three-dimensional case.\cite{Natt_Fis} But it was found later that the combined effect of elastic and electric long-range fields and fluctuations of the walls results in a power-law attraction between the domain walls and, therefore, in the discontinuity of the C-IC transition.\cite{Levanyuk_Laiz.} 

The studies along the third line was also not free of confusion. Bruce, Cowley, and Murray\cite{Bruce} stated that the coupling of the order parameter to any ``long wavelength coordinate'' leaded to the discontinuity of C-IC transitions. The authors mentioned specifically the electrical polarization and the elastic strain, the latter via the strain dependence of the coefficient of the Lifshitz invariant. Bak and Timonen\cite{Bak} criticized this result arguing that taking into account spatial changes of the additional coordinate one results in, simply, a renormalization of the coefficients of the thermodynamic potential that does not change the character of the transition (Bruce, Cowley and Murray forced the coordinate to be uniform while allowed space variations of order parameter). Paradoxically, Bak and Timonen referred to the elastic strain as this additional coordinate but their criticism is perfectly valid for any coordinate just excluding the elastic strain that need a special treatment.\cite{Golovko,Larkin} The matter is in the long-range character of the elastic forces in solids according to the fact that to characterize elastic deformations of solids one uses different variables for the spatially homogeneous and spatially inhomogeneous parts of the deformations: six components of the strain tensor for the homogeneous part and three components of the displacement vector for the inhomogeneous one. Bak and Timonen overlooked the uniform part of the deformations while Bruce, Cowley and Murray excluded from consideration the inhomogeneous part. Subsequent works\cite{Golovko,Cowley} overcome this controversy taking into account both types of deformations and showing that the results of Ref.\cite{Bruce} remained qualitatively correct.

If Bruce, Cowley and Murray had referred their theory to a solid with infinite shear modulus their results would be strictly correct. We shall use this possibility to illustrate the origin of the striction-mediated attraction between solitons considered in the present paper. If the shear modulus of a solid is infinite its only possible deformation is an homogeneous dilatation $\epsilon$. Suppose that a finite density $n$ of domain walls, or more generally solitons, is created in the solid. The soliton self-energy depends, naturally, on $\epsilon$. Taking the state without solitons as the non-deformed one, we present the change of the system energy per volume unit as
\begin{align}
F (u) \simeq  (E_0  +  E_1 \epsilon)n +{\widetilde K\over 2}  \epsilon^2,
\end{align}
were the terms in parentheses represent the soliton self-energy and $\widetilde K$ is the bulk modulus of the system. Minimizing $F$ with respect to $u$ one finds that the equilibrium deformation is $\epsilon_{\rm eq}=- n E_1/\widetilde K$. Then, the change in energy becomes into
\begin{align}
F (\epsilon_{\rm eq}) \simeq   E_0 n  -  {E_1^2  \over 2\widetilde K}n^2,
\end{align}
where the second term represents the attraction mentioned before.\cite{nota} Evidently one can not forget that some soliton repulsion exists as well, a repulsion that provides a finite value of the equilibrium soliton density. Below we reproduce this soliton attraction, in cases in which these solitons represent domain walls, taking into account both homogeneous and inhomogeneous deformations as well as the elastic anisotropy of real crystals.

\vspace{-.25cm}
\section{C-IC phase transition}
\vspace{-.25cm}
To describe the C-IC phase transition we shall use the Landau free energy expressed in terms of the order parameter of the ``virtual'' normal-commensurate (N-C) transition. We restrict ourselves to the case of two-components order parameter $\eta =(\rho \cos \varphi,\eta _2=\rho \sin \varphi)$ and an IC modulation along only one crystallographic direction. In a coordinate frame with the $z$ axis along the IC modulation, the Landau thermodynamic potential can be written as (see, e.g., Ref.\cite{Sannikov}) 
\begin{subequations}\label{F}
\begin{align}
&\qquad \qquad \qquad \qquad F=F_1+F_2,
\\ \nonumber \\
%\begin{split}
F_1&= {1 \over v} \int \Bigl\{ 
     \alpha \rho ^2 
    +\beta  \rho ^4 
    +\gamma \rho ^m \cos (m\varphi) 
    -\sigma \rho ^2 \nabla_z \varphi \Bigr.                                 
    \nonumber \\&\; \;\quad \qquad \left.
    +\delta \left[ \rho ^2 \left(\nabla_z \varphi\right)^2+\left(\nabla_z \rho \right)^2 \right]\right\}dv,
\label{F1}%\end{split}
\\ %\nonumber \\  
F_2&= {1 \over v} \int \left\{
      r_{ij} u_{ij} \rho ^2
     +g_{ij} u_{ij} \rho ^2 \nabla_z \varphi                                 
     +{1\over 2} \lambda_{ijkl}u_{ij}u_{kl}\right\}dv.\label{F2}
\end{align}\end{subequations}
Here $v$ is the volume of the system, $\alpha =\alpha _T(T-\Theta )$ is the only temperature-dependent coefficient, $\gamma$ is related with the anisotropy in the order parameter space, and $m\ge 3$ is an integer that depends on the system in question. It is necessary to assume that the coefficients $\beta $ and $\delta$ are positive from the stability conditions for the free energy. In Eq. \eqref{F2} $u_{ij}$ represents the strain tensor describing deformations of the normal phase (summation over double indices is implied). The term $r_{ij}u_{ij}\rho^2$ takes into account the strain dependence of the coefficient $\alpha$, i.e., the striction effect, and the term $g_{ij} u_{ij} \rho ^2 \nabla_z \varphi$ the strain dependence of the coefficient of the Lifshitz invariant. 

Without these dependencies on the strain, the N-IC transition takes place at a temperature $T_i>\Theta $ defined by the condition $\alpha _0 \equiv \alpha _T(T_i-\Theta )=\sigma ^2/(4\delta )$. This N-IC transition is due to the presence of the Lifshitz invariant because if $\sigma $ were zero only a N-C transition should take place at $T=\Theta$. At the N-IC transition temperature $T_i$, the ``wave number'' of the IC structure is $q_0=\sigma /(2\delta )$. Because of the anisotropy in the order parameter space, as the temperature decreases the IC structure becomes a domainlike one and finally, at some temperature $T_{loc}$, the domain walls disappear in the IC-C transition (so-called lock-in transition). 

If we minimize Eq. \eqref{F} with respect to all elastic degrees of freedom, the resulting thermodynamic potential contains nonlocal terms (see below), which makes it difficult to determine the spatial distribution of the order parameter. But due to the specific features of an isotropic system with infinite shear modulus, to minimize Eq. \eqref{F} for such a system results quite simple having in mind already known results. As we shall see, the results obtained for this ``hypothetical'' case can be easily generalized for real ones.
\vspace{-.25cm}
\subsection{Elasticity-induced attractive interaction}
\vspace{-.25cm}
\subsubsection{Isotropic system with infinite shear modulus}
\vspace{-.25cm}
We consider first the case of an isotropic system with infinite shear modulus. Its only possible deformation is an homogeneous dilatation $u$. If we minimize Eq. \eqref{F} with respect to $u$ we obtain 
\begin{align}
F_2=-{1 \over 2 K}
\left[ r^2 \langle \rho^2 \rangle ^2 
+ 2rg \langle \rho^2 \rangle 
  \langle \rho^2 \nabla_z \varphi \rangle  
+ g^2 \langle \rho^2 \nabla_z \varphi \rangle^2  
\right],
\label{no_local_mu}  
\end{align}
where $r_{ij}=r\delta_{ij}$, $g_{ij}=g\delta_{ij}$, $K$ represents the bulk modulus of the system and $\langle \dots \rangle$ means volume average. As we have mentioned, further minimization of the Landau potential is not very easy due to the contribution of those nonlocal terms.

But in this special case in which the shear modulus of the system is infinite there is another way of minimization. Note that the Landau potential Eq. \eqref{F} can be written as 
\begin{align}
F={1 \over v} \int &\Bigl\{ \alpha(u) \rho ^2 
    +\beta  \rho ^4 
    +\gamma \rho ^m \cos (m\varphi) 
    -\sigma(u) \rho ^2 \nabla_z \varphi\Bigr. 
    \nonumber \\ &\left.%\nonumber \\
    +\delta \left[ \rho ^2 \left(\nabla_z \varphi\right)^2+\left(\nabla_z \rho \right)^2 \right]\right\} dv 
    +{K\over2}u^2,
\label{Fu}    
\end{align}
where $\alpha(u) =\alpha + r u$, $\sigma(u) =\sigma + g u$, and $u$ is a variational parameter. Fixing for a moment this parameter one obtains the same equations of equilibrium for the order parameter as those, e.g., in Ref.\cite{Sannikov}. Their solution gives, in particular, the thermodynamic potential close to the C-IC transition as a function of the domain wall density $n$. In the weak anisotropy case (see below) this can be written as\cite{Sannikov} 
\begin{align}
F= F_c(u) + {K\over 2 } u^2  %\nonumber \\
+E(u)n + 4J(u)n \exp\left[-{m p(u) \over 2n}\right],
\label{F_sannikov}\end{align}
where the coefficients of this expression, in terms of the square of the order parameter amplitude $\rho^2(u) = - {\alpha(u)/(2\beta)}$ and the ``wave number'' of the IC structure at the N-IC transition $q_0(u)=\sigma(u)/(2\delta)$, are 
\begin{gather}
p^2(u)={2 \gamma\over \delta}\rho^{m-2}(u),\\
J(u)={4\sigma(u)\rho^2(u)p(u)\over m q_0(u)},\\
E(u)={J(u)\over 2p(u)}\left[2p(u)-\pi q_0(u)\right],\\
F_c(u)= - \beta \rho^4(u).%-{\alpha^2(u)\over 4\beta}.
\end{gather} 

Let us analyze the physical meaning of Eq. \eqref{F_sannikov} for a clamped crystal, i.e., a crystal with a fixed deformation $u$. The first two terms represent the free energy of the C phase ($n=0$). The domain-wall self-energy is obtained from the third one. It is positive in the C phase, vanishes at the transition point, and is negative in the IC phase. The exponential term represents a repulsion between domain walls. Hence, only if the domain wall self-energy is negative the formation of domain walls is energetically favorable. The minimization of Eq. \eqref{F_sannikov} with respect to $n$ for a clamped crystal yields a continuous C-IC transition at 
\begin{align}
\alpha_c(u)=-2\beta \rho_c^2(u)= -2\beta \left[{\pi^2 \sigma^2(u) \over 2^5\gamma \delta}\right]^{2/(m-2)},
\end{align}
in which the domain wall density vanishes as $\ln^{-1} \{[\alpha(u)-\alpha_c(u)]/\alpha_c(u)\}$.

Let us now turn to consider an unclamped crystal. Its deformation is also a degree of freedom and, therefore, we must minimize Eq. \eqref{F_sannikov} with respect to $u$. In the C phase this deformation is $u_c=r \alpha/(2\widetilde K \beta)$, where $\widetilde K = K - r^2/(2\beta)$. In the IC phase there is, in addition, a deformation $\epsilon \equiv u-u_c$ induced by the creation of domain walls. This is small close to the C-IC transition and so in Eq. \eqref{F_sannikov} only lowest-order terms in $\epsilon$ are relevant. 

Further results show that the strain dependence of the repulsion term can be neglected. Therefore, close to the C-IC transition it can be taken as 
\begin{align}
4J(u)n \exp\left[-{ mp(u)\over 2n}\right]
\simeq a_2 n\exp(-N/n),
\end{align}
where $N=m\pi q_0(u_c)/4$ and $a_2=-4\pi \sigma(u_c)\alpha_c(u_c)/(m\beta)$. In contrast, the strain dependence of the domain-wall self-energy is essential close to the C-IC transition. This is such that 
\begin{align}
E(u)\simeq E_0 + E_1\epsilon, 
\end{align} 
where $E_0 =
{\mathcal E} \left[ \alpha(u_c)-\alpha_c(u_c) \right]$, $E_1 = {\mathcal E}\{ r-4g\alpha_c(u_c)/$ $[(m-2)\sigma(u_c)]\}$ with ${\cal E} ={(2-m)\pi \sigma(u_c)/(2m\beta)}$. As a result we have that close to the C-IC transition Eq. \eqref{F_sannikov} can be written as 
\begin{align}
F\simeq& \widetilde F_c 
+{\widetilde K \over 2 }\epsilon ^2
+(E_0 + E_1 \epsilon ) n + 
a_2n\exp ({-N/n }),
\label{pow_epsil}
\end{align}
where $\widetilde F_c = -\alpha^2 K /(4\widetilde K\beta)$. Note that the strain dependence of the repulsion term yields an exponentially small correction to $E_1$.
 
Minimizing Eq. \eqref{pow_epsil} with respect to $\epsilon$ one obtains that $\epsilon_{\rm eq}=-nE_1/\widetilde K$. In consequence, the free energy as a function of the domain-wall density has the form
\begin{align}
F=& \widetilde{F}_c + a_1 n + a_2 n \exp ({-N/n }) - a_3 n^2,
\label{Fq}
\end{align}
where $a_1=E_0$, $a_2$ has been defined before and $a_3=E_1^2/(2\widetilde K)$. The term $-a_3n^2$ describes an attractive interaction between domain walls and leads to a discontinuity of the C-IC transition as we will show.%\cite{comment} 

Let us mention that within the constant-amplitude approximation used, the equation  $K u + r \langle \rho^2 \rangle =0 $ which follows from Eq. \eqref{F} in absence of external forces taking $g=0$ is not satisfied. In order to have it satisfied, higher order corrections to $\rho^2$ in the so-called small anisotropy parameter ($\varepsilon_m$ in the following) must be taken into account. However, these corrections lead to higher order terms in $\varepsilon_m$ in the thermodynamic potential\cite{comment} and therefore are not important for the attraction.

The two types of coupling between the order parameter and the strain considered in Eq. \eqref{F} contribute to this attractive interaction. The ratio between these two contributions in the coefficient $a_3$ is 
\begin{align}
%{G \over r} = 
{-4g \over (m-2)r } {\alpha_c(u_c) \over \sigma(u_c)} =
{m\pi^2 \over 2^3(m-2)}{g \over r } {q_0(u_c) \over \varepsilon_m},
\label{G_r}\end{align}
where the parameter $\varepsilon_m =-(m\pi^2/2^4)[ \alpha_0 /\alpha_c(u_c) ]$ defines the condition of weak anisotropy in the order parameter space.\cite{Sannikov} Eq. \eqref{F_sannikov} is valid by virtue of the smallness of this parameter. 

The typical periods of the IC structures are larger than the atomic distances $d_{at}$: usually they are such that $q_0(u_c)\sim 10^{-2}d_{at}^{-1}$. The ratio $g/r $ can be roughly estimated as $d_{at}$ from dimensional arguments. Therefore, the relative importance of the two elasticity-mediated interactions is determined by the ratio between two small parameters $q_0(u_c)d_{at}/\varepsilon_m$. As a result, the striction-mediated interaction is the most important one if $10^{-2}<\varepsilon_m<1$. Recall that there are also systems without Lifshitz invariant (type II) where the striction-mediated attraction is unrivaled.

For $m=3,4,6$ we have
\begin{gather}
\varepsilon_3={4 \delta \over \beta }\left[{\gamma \over \sigma(u_c)}\right]^2,\qquad
\varepsilon_4={\gamma \over \beta },\qquad
\varepsilon_6={ \sigma(u_c) \over 2 \beta }\left( {\gamma\over \delta}\right)^{1/2}.
\end{gather}
The smallness of $q_0(u_c)=\sigma^2(u_c)/(4\delta)$, in atomic units, implies the smallness of $\sigma(u_c)$ since $\delta$, which gives the curvature at the minimum of the soft branch at the N-IC transition, has usually normal values. Therefore, the condition of weak anisotropy is easily fulfilled for $m=6$ even if all other coefficients have normal atomic values. However, it is not the case for $m=4,3$ which need, in addition, a special smallness of $\gamma$.
\vspace{-.25cm}
\subsubsection{Real systems}
\vspace{-.25cm}
We shall show that the free energy of real systems in terms of the domain-wall density has the same form that Eq. \eqref{Fq} close to the C-IC transition. To minimize the thermodynamic potential Eq. \eqref{F} with respect to the elastic degrees of freedom we first divide $u_{ij}$ into spatially homogeneous and inhomogeneous parts. The minimization should be carried out separately for these two parts because they represent degrees of freedom of the system independent one of each other.\cite{Larkin} The above-mentioned division can be presented as 
\begin{align}
u_{ij}=u_{ij}^{(0)}
 +{i\over 2} \sum_{{\bm k}\neq 0} 
 \left[k_iu_j({\bm k})+k_ju_i({\bm k})\right]e^{i{\bm k \cdot \bm r}},
\end{align}
where $u_i({\bm k})$ are the Fourier components of the displacement vector. Because only variations along the $z$ axis are allowed we have that ${\bm k} = (0,0,k_z)$. 

Let us consider first the case in which the striction is the only coupling between the order parameter and the strain, i.e., we take $g_{ij}=0$ in Eq. \eqref{F2}. As a result of the minimization of Eq. \eqref{F2} with respect to the strain one obtains 
\begin{align}
F_2=
-{1\over 2}r_{ij}r_{kl}
\lambda_{ijkl}^{-1}f^2_0
-{r_{33}^{2}\over 2\lambda_{3333}} \sum \limits _{k_z\neq 0}
f_{k_z}f_{-k_z}
\label{} 
\end{align}
in the Fourier space, where $f_{k_z}$ represent the Fourier components of the function 
%\begin{align}
$f\equiv \rho^2$ and $\lambda_{ijkl}^{-1}$ is the $ijkl$ component of the inverse of the tensor $\lambda_{ijkl}$. %= \sum_{k_z}f_{k_z}e^{ik_zz}$. 
%\label{} 
%\end{align}
In the real space this is
\begin{align}
F_2=
-{1\over 2}\left[r_{ij}r_{kl}
\lambda_{ijkl}^{-1}-{r_{33}^{2}\over \lambda_{3333}} \right]\langle\rho^2\rangle^2
-{r_{33}^{2}\over 2\lambda_{3333}} \langle \rho^{4} \rangle.
\label{no_local} 
\end{align}
Note that the first term of this functional is formally identical to Eq. \eqref{no_local_mu} if, for that case of infinite shear modulus, we take $g=0$. Note also that the contribution of the second term to the thermodynamic potential can be understood as a renormalization of the coefficient $\beta$ in Eq. \eqref{F1}. As a consequence, the free energy of a real crystal in terms of their domain-wall density has the form Eq. \eqref{Fq} with the renormalized constants
\begin{gather}
\beta \quad \to \quad \beta - r_{33}^2/(2\lambda_{3333}),
\\ %\nonumber \\ 
r^{2}/K \quad \to \quad r_{ij}r_{kl}\lambda_{ijkl}^{-1}- r_{33}^2/\lambda_{3333}.
\end{gather}

If both striction and Lifshitz invariant mediated couplings are taken into account, in addition of these renormalizations we have 
\begin{gather}
g^{2}/K \quad \to \quad g_{ij}g_{kl}\lambda_{ijkl}^{-1}- g_{33}^2/\lambda_{3333},
\\ %\nonumber \\
rg/K \quad \to \quad r_{ij}g_{kl}\lambda_{ijkl}^{-1}- r_{33}g_{33}/\lambda_{3333}.
\end{gather}
In this case, after the minimization over the strain one obtains also terms that renormalize the coefficients of the invariants $\rho^4\nabla_z\varphi$ and $\rho^4(\nabla_z\varphi)^2$. These invariants were not considered previously in Eq. \eqref{F} because their role is not essential. Therefore these contributions can be neglected. 

Let us discuss an isotropic case in which the shear modulus $\mu$ of the system is finite. The elastic moduli of such a system are $\lambda_{ijkl}=[K-(2\mu/3)]\delta_{ij}\delta_{kl} + \mu(\delta_{ik}\delta_{jl} +\delta_{il}\delta_{jk} ) $. One can see that the above mentioned renormalization of the coefficients in Eq. \eqref{Fq} results in
\begin{gather}
\beta \quad\to\quad
\beta-{r^2\over 2\left(K + {4\over 3}\mu\right)},
\\ %\nonumber \\
r^{2}/K \quad \to \quad 
{4r^{2}\mu \over 3 K\left( K+{4\over 3} \mu \right)},
\\ %\nonumber \\
g^{2}/K \quad \to \quad 
{4g^{2}\mu \over 3 K\left( K+{4\over 3} \mu \right)}.
%\\ \nonumber \\
%rg/K \quad \to \quad 
%{4rg\mu \over 3 K\left( K+{4\over 3} \mu \right)}.
\end{gather}
If the striction-mediated contribution to the attraction term of Eq. \eqref{Fq} is neglected, i.e., one takes $r=0$, the resulting attraction term coincides with the previously reported in Refs.\cite{Golovko,Sannikov}.
\vspace{-.25cm}
\subsubsection{Other contributions to the attractive interaction}
\vspace{-.25cm}
For IC ferroelectrics thermal fluctuations of domain walls induce macroscopic electric fields which generate a van der Waals-like attraction between them.\cite{Levanyuk_Laiz.} Its contribution to free energy of the system is  
\begin{align}
F_{\rm vdW}=
\begin{cases}
-{\displaystyle 3 \over \displaystyle 32 \pi}
 {\displaystyle T\over\displaystyle h_{D}^{2}l}
 \ln {\displaystyle h_{D}\over \displaystyle l} & (l<h_{D}),\\
\\  
-{\displaystyle 1 \over \displaystyle 16 \pi} 
 {\displaystyle T \over \displaystyle l^{3}} & 
 (l>h_{D}),\\
\end{cases}\label{}
\end{align}
where $l$ is distance between domain walls, $h_{D} \simeq d_{at} [T_{0}/(\Theta -T)]^{3/2}$ and it has been taken $m=6$ ($\rm K_2SeO_4$-type crystals). 

Let us estimate, at the C-IC transition, the ratio between this contribution and the elasticity-induced one considering the striction-mediated part only (the striction-mediated attraction and that due to the Lifshitz invariant are expected to be of the same order of magnitude for $\rm K_2 Se O_4$ as we shall see below). 

From the last term of Eq. \eqref{Fq}, the elasticity-induced attraction between domain walls at the C-IC transition point is
\begin{align}
F_{\rm el}%\simeq
%- {\pi ^4 \over 2}{r^2 \over K }{4\mu \over 3\widetilde{K}}{\delta^2 q_0^4 \over  \beta ' \beta ''} \widetilde{q}_{loc}^2
\simeq-(2\pi)^2{\Delta K \over K}
{T_{0}d_{at}  \over \lambda _0^2 \lambda_{loc}^2},
\label{}
\end{align}
where $\Delta K/K\sim10^{-1}\div 10^{-2}$ is the relative change of the bulk modulus between the normal and the incommensurate phases, $d_{at}$ is the interatomic distance, $\lambda_0=2\pi / q_0(u_c) \sim 10^2d_{at}$ is the typical wavelength of the IC phases at the N-IC transition, $\lambda_{loc} \gtrsim \lambda _0$ is the typical wavelength of the IC phases at the C-IC one, and $T_{0}$ is equal to $T_{at}\sim 10^4\div 10 ^5 \rm K$ for displacive systems and to $T_{loc}$ for order-disorder ones. Thus, the above mentioned ratio is found to be
\begin{align}
{F_{\rm vdW}\over F_{\rm el}}\simeq
\begin{cases}
(10^5 \div 10) 
{\displaystyle T_{loc}\left(\Theta - T_{loc}\right)^3
 \over \displaystyle T_0^4}
& (l_{loc}<h_{D}),\\
\\  
(10 \div 10^{-1})  
{\displaystyle T_{loc} \over \displaystyle T_{0} }
& (l_{loc}>h_{D}).\\
\end{cases}
\label{}
\end{align}
Taking into account that typically $\Theta - T_{loc} \simeq T_{loc}$, one finds that the contribution of van der Waals-like attraction would be greater than striction-mediated one for order-disorder systems ($T_0 \sim T_{loc}$), but it turns out to be as much as one order-of-magnitude smaller than the striction-mediated contribution for displacive systems ($T_0\sim T_{at}$).

We mention that there is as well a universal fluctuation-induced attraction between domain walls operative by virtue of the solid-state elasticity.\cite{Levanyuk_Laiz.} However its effects are orders of magnitude smaller than the elasticity-induced interaction considered in this paper both for order-disorder and displacive systems. 
\vspace{-.25cm}
\section{Thermal anomalies close to the C-IC transition}
\vspace{-.25cm}
\subsection{Theoretical formulaes}
\vspace{-.25cm}
In this section we deduce formulas for thermal anomalies that a free energy like Eq. \eqref{Fq} describes. The procedure is similar to that exposed in Ref.\cite{Martin-Ollala} but keeping in mind the different form of the attraction term. 

The coefficient $a_1$ of Eq. \eqref{Fq} can be expressed as $a_1=a_{1T}(T-T_c)$ where $T_c = \Theta 
+ (\widetilde K/K) [\alpha_c(u_c)/\alpha_T]$ is the temperature at which the domain-wall self-energy changes its sign. However, due to the domain-wall attraction, $T_c$ does not coincide with the C-IC transition temperature that will be determined below.

The minima (stable or metastable) of the free energy Eq. \eqref{Fq} are obtained at certain values of the domain wall density, say, $n_e$, such that the conditions
\begin{gather}
\left.{\partial F \over \partial n} \right|_{n_e}=a_1+a_2(1+N/n_e)e^{-N/n_e}-2a_3n_e=0,  
\label{e30}
\\
%\nonumber \\
\left. \partial^2F \over \partial n^2\right| _{n_e\not =0}=
\frac{2a_3n_e\left( N^2-Nn_e-n_e^2\right) -a_1N^2}
{n_e^2\left(N+n_e\right) }  
>0,
\end{gather}
are fulfilled. Note that at $T_1=T_c-[2a_3n_e/(a_{1T}N^2)]\left( N^2-Nn_e-n_e^2\right)$
it is obtained $\left( \partial ^2F /\partial n^2\right) _{n_e\not=0}=0$
and, therefore, it represents the limit of undercooling of the IC phase: below this temperature any minimum of the free energy (stable or metastable) is an state without domain walls. 

From Eq. \eqref{e30} and the condition of continuity of the free energy at the C-IC transition point, i.e., $F=\widetilde{F}_c$, we obtain that
\begin{gather}
T_{loc}=T_c-{a_3n_{loc}(N-n_{loc})/(a_{1T}N)},  
\\ %\nonumber \\
n_{loc}=N\ln^{-1} \left[{a_2N /(a_3n_{loc}^2) }\right],
\label{e32}
\end{gather}
give, respectively, the C-IC transition temperature and the density of domain walls at the C-IC transition point. As we see, as long as $a_3$ is not zero the C-IC transition temperature does not coincide with $T_c$ and at this temperature the domain-wall density is not zero.

The latent heat of the C-IC phase transition is 
\begin{align}
Q=T_{loc}a_{1T}n_{loc},  
\label{e33}
\end{align}
and the specific heat of the IC phase depends on the temperature as 
\begin{align}
C_p = \frac{C_1}{T-T_1},  
\label{e36}\end{align}
with a Curie constant
\begin{align}
C_1=Ta_{1T}(n_e/N)^2  (N+n_e).
\label{C_1}
\end{align}
Note that $C_p$ diverges at the limit of undercooling of the IC phase. We mention that within a frame of a continuous C-IC transition the anomaly of the specific heat has the same form.\cite{Sannikov_Golovko} 

For IC ferroelectrics, e.g., $\rm K_2SeO_4$-type crystals, the electric susceptibility in the IC phase is also interesting. The attraction term in Eq. \eqref{Fq} is proportional to the square of the domain walls concentration. Therefore, it does not contribute to the electric susceptibility since the electric field changes the distance between domain walls but not its concentration. Thus, from previous works (see e.g., Refs.\cite{Sannikov,Martin-Ollala}) and from Eq. \eqref{e30} we have approximately (taking $n_e\simeq n_{loc}$) the electric susceptibility as  
\begin{equation}
\chi \simeq\frac{C_2}{T-T_2},  
\label{e37a}
\end{equation}
where 
\begin{gather}
C_2=\frac{P_s^2}{a_{1T}}(N+n_{loc}), 
\label{C_2}
\\ %\nonumber \\
T_2=T_c-\frac{2a_3n_{loc}}{a_{1T}}< T_1,
\end{gather}
with $P_s$ being the polarization within the domains. 

Note that at $T=T_1$, the temperature where the specific heat formally diverges, the electric susceptibility remains finite. This might explain the experimental observation reported in Ref.\cite{Martin-Ollala}, according to which the maximum of specific heat occurs at a higher temperature than the maximum of the dielectric constant. 

It should be mentioned that the compressibility and the thermal-expansion coefficient also exhibit an anomaly similar to that of the specific heat. In other words, the compressibility of the crystal in the IC phase close to the lock-in transition can be quite high. However this is valid only for the static compressibility because the ``softness'' is associated with changes in the domain-wall lattice under the influence of the external pressure. Since the domain-wall processes are very slow one has to use very low frequencies measurements to probe this ``softness.''

\subsection{Comparison with experiments: K$_{\text{2}}$SeO$_{\text{4}}$ and Rb$_{\text{2}}$ZnCl$_{\text{4}}$}

There is an extensive literature devoted to the experimental study of the anomalies associated with C-IC transitions (see, e.g., Ref.\cite{IC_phases_in}). Those which are focused on $\rm K_2SeO_4$ result of special interest for us due to the following reasons. First, it is known that the free-energy expansion results in the form of Eq. \eqref{F} (with only one temperature-dependent coefficient) is valid to describe both the C-IC transition and the N-IC one.\cite{Sannikov_Golovko} It allows us, in principle, to calculate the coefficients of the free energy Eq. \eqref{Fq} independently on data of the C-IC transition, i.e., the coefficients can be determined from the experimental data about the N-IC transition. Second, so far reported data permit to do it. In fact this valuable work was done previously by Sannikov and Golovko.\cite{Sannikov_Golovko} All coefficients of the functional Eq. \eqref{F} were consistently determined with the exception of $g_{ij}$, which needs more complete studies on the pressure dependence of the N-IC phase transition. According to the existing studies\cite{Press} one can see that the above estimation $g/r \sim d_{at}$ is correct by an order of magnitude. Using the value $\varepsilon_6 \sim 10^{-2}$ given in Ref.\cite{Sannikov_Golovko} one sees from Eq. \eqref{G_r} that contributions of the two elasticity-induced mechanisms are comparable for $\rm K_2SeO_4$. Still in what follows we shall consider the striction-mediated interaction as the only one mechanism responsible of the attraction between domain walls, discussing at the end possible consequences of this assumption. Thus, our comparison between theory and experiments for $\rm K_2SeO_4$ can be made with no fitted parameters and we know no other example where this is possible.  

It has been already mentioned that in the interpretation of the experimental data an important characteristic of the ``domain-wall systems'' must be taken into account: they relax to equilibrium so slowly that it is impossible to observe their equilibrium states. It is evidenced in the hysteresis phenomena typical of the IC phases.\cite{hysteresis} It implies that the observed thermal anomalies really correspond to smeared phase transitions predicted theoretically for systems in their equilibrium states. Hence, the aim of any comparison between theory and experiments must be to see if the experimental data can be reasonably well explained as a result of the smearing of the equilibrium (theoretical) phase transitions. This fact resolves the apparent contradiction between reported experiments so far, which show a great number of continuous C-IC anomalies,\cite{pie} and the theory, which predicts discontinuous C-IC transitions. 
It motivates also a preferential comparison between integral characteristics such as the enthalpy of the transition. 

For $\rm K_2SeO_4$ whose non-zero elastic moduli are\cite{Sannikov_Golovko}
%\begin{subequations}
\begin{gather}
\lambda_{ii}=\left\{5.3,5.0,3.6,0.48,1.5,1.6 \right\},
\nonumber \\ \\
\lambda_{12,13,23}=\left\{1.7,1.5,2.0\right\},\nonumber 
\label{} 
\end{gather}
%\end{subequations}
in units of $10^{11}~\rm dyn\cdot cm^{-2} $, 
\begin{align}
r_{i}/\alpha _0 =\left\{0.7,2,-8,0,0,0\right\}\cdot 10^{3},
\label{} 
\end{align}
(abbreviated notation is used and the index $i$ runs the values $1,2,\dots 6$) and $\beta \widetilde K / (K \alpha_0) = 1.3 \cdot 10^{-4}~\rm dyn^{-1} \cdot cm^2$, we find that $a_3N/a_2 \simeq 10^{-2}$. Then Eq. \eqref{e32} gives $n_{loc}\sim 0.1N$ and therefore the change of the ``wave naumber'' of the IC modulation would be such that $q_0/q_{loc}\sim 2$ ($q_{loc}=2\pi n_{loc}/m$). From reported data in Ref.\cite{Iizumi}, the experimental value of this ratio is $q_0/q_{loc}\sim 3.5$. The difference between both values can be commented in the following way. First note that the separation between the nearest points to the C-IC discontinuity in the experimental curve of $q(T)$ (Fig. 4 in Ref.\cite{Iizumi}) is about $4~\rm K$. But as it can be seen in Ref.\cite{Chaudhuri}, all the observable C-IC thermal anomaly occurs into a region of about $\rm 1~K$. Therefore, one cannot be sure that the experimental value really corresponds to the wave vector at the lock-in transition. It might be even more important the fact that the observed wave vector of the IC modulation never corresponds to its equilibrium value close to the C-IC transition because the relaxation times of the system are much greater than the measurement times. Thus, what is really meant by the experimental ratio $q_0/q_{loc}$ remains somewhat undetermined. Second, it is quite possible that striction and Lifshitz invariant mediated contributions to $a_3$ compensate each other. Indeed both transition temperatures and $q_0$ diminish with the pressure,\cite{Press} a behavior that for an isotropic case evidence opposite signs of $r$ and $g$.  

To the best of our knowledge, the most detailed study about the thermal anomalies of $\rm K_2SeO_4$ was made in Ref.\cite{Ishibashi}. Unfortunately, the discussion of the reported data was based on a theory\cite{Ishibashi,Golovko80} in which the prediction of the discontinuity of the C-IC transition is erroneous because no discontinuity is obtainable if the considered model is treated consistently.\cite{Sannikov} We proceed to illustrate that the thermal anomaly for the specific heat of $\rm K_2SeO_4$ can be reasonably well understood as the smearing of a transition which is discontinuous mainly due to the striction-mediated attraction between domain walls.

For $\rm K_2SeO_4$ according to Ref.\cite{Sannikov_Golovko} it is obtained the value $a_{1T}\simeq 0.05\div 0.5~\rm J\cdot mol^{-1}\cdot K^{-1}$ and therefore, from Eq.\eqref{e33} with $T_{loc} = 95~\rm K$, the latent heat of the lock-in transition is $Q \sim 0.6 \div 4~\rm  J \cdot mol^{-1}$. This latent heat is enough to be observed experimentally if the transition (equilibrium) were really observed, but its smearing is such that the thermal anomaly observed appears as continuous.\cite {Chaudhuri} In order to estimate the contribution of the anomaly of $C_p$ to the enthalpy of the C-IC transition we calculate the difference between the C-IC transition temperature and the temperature at the limit of undercooling of the IC phase (the specific heat diverges at this limit). It is obtained a temperature difference of about $0.5~\rm K$. From Eq.(\ref{C_1}) the Curie constant for the specific heat at the lock-in temperature is $C_1 \sim 0.1 \div 0.03 $ $\rm J \cdot mol^{-1}$. Bearing in mind that all the C-IC anomaly occurs into a region of about $1~\rm K$, the contribution of the Curie anomaly to the enthalpy results one order-of-magnitude lower than that of the latent heat. Therefore, theoretical enthalpy of the C-IC transition is in reasonable agreement with their experimental value\cite {Chaudhuri} $1.1~\rm J \cdot mol^{-1}$, and it can be said that almost all contribution to the thermal anomaly comes from the smearing of the latent heat. 

Another fairly-well studied compound is $\rm Rb_2ZnCl_4$. However we do not dispose of the data necessary to calculate the coefficients of the free energy in the same way as the before for $\rm K_2SeO_4$. But we can use the experimental value of the Curie constant $C_2\simeq 5\;\rm K$, which seems not to vary substantially for different samples (unlike to the Curie constant $C_1$, see e.g. Ref.\cite{Martin-Ollala}), as well as the data for $P_s =0.12 \; \rm \mu C\cdot cm^{-2}$ and $n_{loc} \simeq 0.1 \div 0.03N$ (see Refs.\cite{Martin-Ollala,Hamano-Novotna}) to estimate the values of the latent heat and the Curie constant $C_1$. Thus, from Eqs. \eqref{C_2}, \eqref{C_1} and \eqref{e33} with $T_{loc}\simeq 195$ $ \rm K$, it is obtained $Q=7\div2$ $ \rm J\cdot mol^{-1}$ and $C_1=0.8 \div 0.06$ $ \rm J \cdot mol^{-1}$. These estimations are in a reasonable agreement with the experimental values reported in Refs.\cite{Martin-Ollala,Atake} of $Q=6\div2$ $ \rm J\cdot mol^{-1}$ and $C_1=1\div0.1$ $ \rm J\cdot mol^{-1}$. 
\vspace{-.25cm}
\section{Conclusions}
\vspace{-.25cm}
We have shown that the most universal interaction between the order parameter and the elastic strain (striction effect) leads to a long-range attraction between solitons. As a consequence, any C-IC transition in a solid should be discontinuous. This striction-mediated attraction has been revealed in detail for a class of C-IC transition that permits a clear and consistent analysis. The obtained results have been compared with experiments showing that the thermal anomalies in $\rm K_2SeO_4$ associated to the C-IC transition can be well understood as a result of the smearing of this transition.  

We mention that we have considered the interaction between the domain walls in the bulk, neglecting the surface effects. The relaxation of the internal stresses and possible changes of the order-parameter distribution close to the surface can modify this attraction. The role of the striction effect in the surface properties of the IC phases deserves a special investigation. It seems promising to study the role of the striction-mediated attraction in other systems, especially in the cases where continuous C-IC transitions are supposed for symmetries not admitting the Lifshitz invariant. 

\vspace{-.25cm}
\section{Acknowledgments}
\vspace{-.25cm}
S.A.M. was supported from the Russian Fund for Fundamental Research (Grants 00-02-16823 and 00-02-17746). A.C.: thanks to L.S. Froufe for useful discussions; special thanks to D. Sanchez for her help and support.

\end{multicols}

\end{document}